\documentclass[12pt,a4paper]{article}
\usepackage{amsmath,amssymb,cite,slashed,tabularx}
\usepackage{subfigure}
\usepackage{graphicx, color}

\allowdisplaybreaks
\newcommand{\TeV}{\,{\rm TeV}}

\makeatletter
\newcommand{\@authornote}[2]{{\def\thefootnote{\fnsymbol{footnote}}\setcounter{footnote}{#1}#2\setcounter{footnote}{0}}}
\newcommand{\authornotemark}[1]{\@authornote#1{\addtocounter{footnote}{-1}\footnotemark}}
\newcommand{\authornotetext}[2]{\@authornote#1{\footnotetext{#2}}}
\makeatother

\usepackage{caption}
\newcommand{\non}{\nonumber \\ }

\def\beq#1\eeq{\begin{align}#1\end{align}}
\newcommand{\kk}{\ensuremath{K^0\textrm{--}\overline{K}{}^0}}

\definecolor{BlueViolet}{rgb}{0.2, 0.00, 0.7}
\definecolor{Blue}{rgb}{0.15, 0.00, 0.9}
\usepackage[colorlinks=true,linkcolor=Blue,citecolor=Blue,urlcolor=BlueViolet]{hyperref}

\begin{document}

\renewcommand{\thefootnote}{\fnsymbol{footnote}} 
\begin{titlepage}

\begin{center}

\hfill KEK--TH--1950\\
\hfill TTP16--057\\
\hfill December 2016

\vskip .75in

{\Large\bf
Revisiting Kaon Physics in General $\boldsymbol{Z}$ Scenario
}

\vskip .75in

{\large\bf
  Motoi Endo$^{\rm (a,b)}$, 
  Teppei Kitahara$^{\rm (c,d)}$,
  Satoshi Mishima$^{\rm (a)}$,\\ 
\vspace{.2cm}}
{\large  and}
{\large\bf  Kei Yamamoto$^{\rm (a)}$
}

\vskip 0.25in

$^{\rm (a)}${\it
Theory Center, IPNS, KEK, Tsukuba, Ibaraki 305-0801, Japan}

\vskip 0.1in

$^{\rm (b)}${\it
The Graduate University of Advanced Studies (Sokendai),\\
Tsukuba, Ibaraki 305-0801, Japan}

\vskip 0.1in

$^{\rm (c)}${\it 
Institute for Theoretical Particle Physics (TTP), Karlsruhe Institute of Technology, Engesserstra{\ss}e 7, D-76128 Karlsruhe, Germany}
 
\vskip 0.1in
 
$^{\rm (d)}${\it
Institute for Nuclear Physics (IKP), Karlsruhe Institute of
Technology, Hermann-von-Helmholtz-Platz 1, D-76344
Eggenstein-Leopoldshafen, Germany} 

\end{center}

\vskip .5in

\begin{abstract}
New physics contributions to the $Z$ penguin are revisited in the light of the recently-reported discrepancy of the direct CP violation in $K\to\pi\pi$.
Interference effects between the standard model and new physics contributions to $\Delta S = 2$ observables are taken into account.
Although the effects are overlooked in the literature, they make experimental bounds significantly severer. 
It is shown that the new physics contributions must be tuned to enhance $\mathcal{B}(K_L \to \pi^{0} \nu \bar{\nu})$, if the discrepancy of the direct CP violation is explained with satisfying the experimental constraints.
The branching ratio can be as large as $6 \times 10^{-10}$ when the contributions are tuned at the 10\,\% level.
\end{abstract}
\end{titlepage}

\setcounter{page}{1}
\renewcommand{\thefootnote}{\#\arabic{footnote}}
\setcounter{footnote}{0}

\section{Introduction}
\setcounter{equation}{0}
\label{sec:intro}

A deviation of the standard model (SM) prediction from the experimental result is recently reported in the direct CP violation of the $K\to\pi\pi$ decays, which is called $\epsilon'$.
The latest lattice calculations of the hadron matrix elements significantly reduced the theoretical uncertainty \cite{Blum:2011ng, Blum:2012uk, Blum:2015ywa, Bai:2015nea} and yield~\cite{Buras:2015yba,Kitahara:2016nld}
\begin{align}
 \left(\frac{\epsilon'}{\epsilon}\right)_{\rm SM} = 
 \left\{
 \begin{array}{ll}
 (1.38 \pm 6.90) \times 10^{-4}, & [\mbox{RBC-UKQCD}] \\
 (1.9 \pm 4.5) \times 10^{-4}, & [\mbox{Buras et al.}] \\
 (1.06 \pm 5.07) \times 10^{-4}. & [\mbox{Kitahara et al.}]
 \end{array}
 \right.
\end{align}
They are lower than the experimental result~\cite{Batley:2002gn,AlaviHarati:2002ye,Abouzaid:2010ny,Olive:2016xmw},
\begin{align}
 \left(\frac{\epsilon'}{\epsilon}\right)_{\rm exp} = (16.6 \pm 2.3) \times 10^{-4}.
   \label{eq:epp_ex}
\end{align}
The deviations correspond to the $2.8$--$2.9\sigma$ level.

Several new physics (NP) models have been explored to explain the discrepancy~\cite{Buras:2014sba,Blanke:2015wba,Buras:2015yca,Buras:2015kwd,Buras:2015jaq,Buras:2016dxz,Tanimoto:2016yfy,Kitahara:2016otd,Endo:2016aws,Bobeth:2016llm,Cirigliano:2016yhc}.
In the literature, electroweak penguin contributions to $\epsilon' / \epsilon$ have been studied.\footnote{
QCD penguin contributions, e.g., through Kaluza-Klein gluons, have also been considered~\cite{Buras:2014sba}. }
In particular, the $Z$ penguin contributions have been studied in detail~\cite{Buras:2012jb,Buras:2014sba,Buras:2015jaq,Buras:2015yca}. 
The decay, $s\to dq\bar q$ ($q=u,d$), proceeds by intermediating the $Z$ boson, and its flavor-changing ($s$--$d$) interaction is enhanced by NP.
Then, the branching ratios of $K \to \pi\nu\bar\nu$ are likely to be deviated from the SM predictions once the $\epsilon'/\epsilon$ discrepancy is explained. 
This is because the $Z$ boson couples to the neutrinos as well as the up and down quarks. 
They could be a signal to test the scenario. 

Such a signal is constrained by the indirect CP violation of the $K$ mesons.
The flavor-changing $Z$ couplings affect the indirect CP violation via the so-called double penguin diagrams; the $Z$ boson intermediates the transition, both of whose couplings are provided by the flavor-changing $Z$ couplings.
Such a contribution is enhanced when there are both the left- and right-handed couplings because of the chiral enhancement of the hadron matrix elements. 
This is stressed by Ref.~\cite{Buras:2015jaq} in the context of the $Z'$-exchange scenario. 
In the $Z$-boson case, since the left-handed coupling is installed by the SM, the right-handed coupling must be constrained even without NP contributions to the left-handed one. 
Such interference contributions between the NP and the SM are overlooked in Refs.~\cite{Buras:2012jb,Buras:2014sba,Buras:2015jaq,Buras:2015yca} \cite{Buras}. 
Therefore, the parameter regions allowed by the indirect CP violation will change significantly. 
In this letter, we revisit the $Z$-boson scenario.\footnote{
In this letter, we focus on the $s$--$d$ transitions. 
The $\Delta F=2$ transitions such as $\Delta m_B$ generally involve the interference contributions.
}
It will be shown that the NP contributions to the right-handed $s$--$d$ coupling are tightly constrained due to the interference, and thus, the branching ratio of $K_L \to \pi^0\nu\bar\nu$ is likely to be smaller than the SM predictions if the $\epsilon'/\epsilon$ discrepancy is explained.
We will discuss that NP parameters are necessarily tuned to enhance the ratio. 
A degree of the parameter tuning will be investigated to estimate how large $\mathcal{B}(K_{L} \to \pi^{0} \nu \bar{\nu})$ and $\mathcal{B}(K^{+} \to \pi^{+} \nu \bar{\nu})$ can become.

\section{$\boldsymbol{Z}$-penguin observables}
\setcounter{equation}{0}

In this section, we briefly review the $Z$-penguin contributions to $\Delta S = 2$ and $\Delta S = 1$ processes in the general $Z$ scenario.
Above the electroweak symmetry breaking scale, NP particles generate Wilson coefficients of the (dimension-6) effective operators,
\begin{align}
  \mathcal{O}_L &= i(H^\dagger \overleftrightarrow{D_\mu} H) (\overline{Q}_L \gamma^\mu Q'_L), \\
  \mathcal{O}_R &= i(H^\dagger \overleftrightarrow{D_\mu} H) (\overline{d}_R \gamma^\mu d'_R), \\
  \mathcal{O}^{(3)}_L &= i(H^\dagger \sigma^a\overleftrightarrow{D_\mu} H) (\overline{Q}_L \gamma^\mu \sigma^a Q'_L).
  \label{eq:effective}
\end{align}
They are gauge invariant under the SM gauge transformations.
In this letter, we focus on the operators, $\mathcal{O}_L$ and $\mathcal{O}_R$, to demonstrate the impact of the interference between the SM and NP contributions.\footnote{
  A similar discussion as follows is expected to hold for the effective operator $\mathcal{O}^{(3)}_L$.
}
After the electroweak symmetry breaking, they provide the flavor-changing ($s$--$d$) $Z$ interactions,
\beq
  \mathcal{L} =& \Delta_L^{\rm NP} \left[ Z_{\mu} + \frac{1}{m_Z} \partial_{\mu} G^0
  - \frac{ig}{ 2 m_W m_Z } G^- \overleftrightarrow{\partial_{\mu}} G^+
  - \frac{g}{m_Z} \left( W_{\mu}^- G^+  + W_{\mu}^+ G^- \right) + \dots \right]
  (\overline{s} \gamma^{\mu} P_L d )\non
  &  + (L \leftrightarrow R) + \textrm{H.c.},
  \label{eq:NPvertex}
\eeq
where the first term in the bracket is the $Z$-boson interaction, while the others are those of the Nambu-Goldstone boson, and we omitted the irrelevant terms for the interference effects.
Here, the Wilson coefficients of $\mathcal{O}_L$ and $\mathcal{O}_R$ are normalized by the flavor-changing $Z$ interactions. 
In the following, we omit the subscript ``NP'' in $\Delta_L^{\textrm{NP}}$ and $\Delta_R^{\textrm{NP}}$ for simplicity.

\subsection{$\boldsymbol{\epsilon_K} $ and $\boldsymbol{\Delta m_K}$}

\begin{figure}[t]
\begin{center}
\subfigure[]
{
\includegraphics[width=0.15\textwidth, bb = 0 0 842 762]{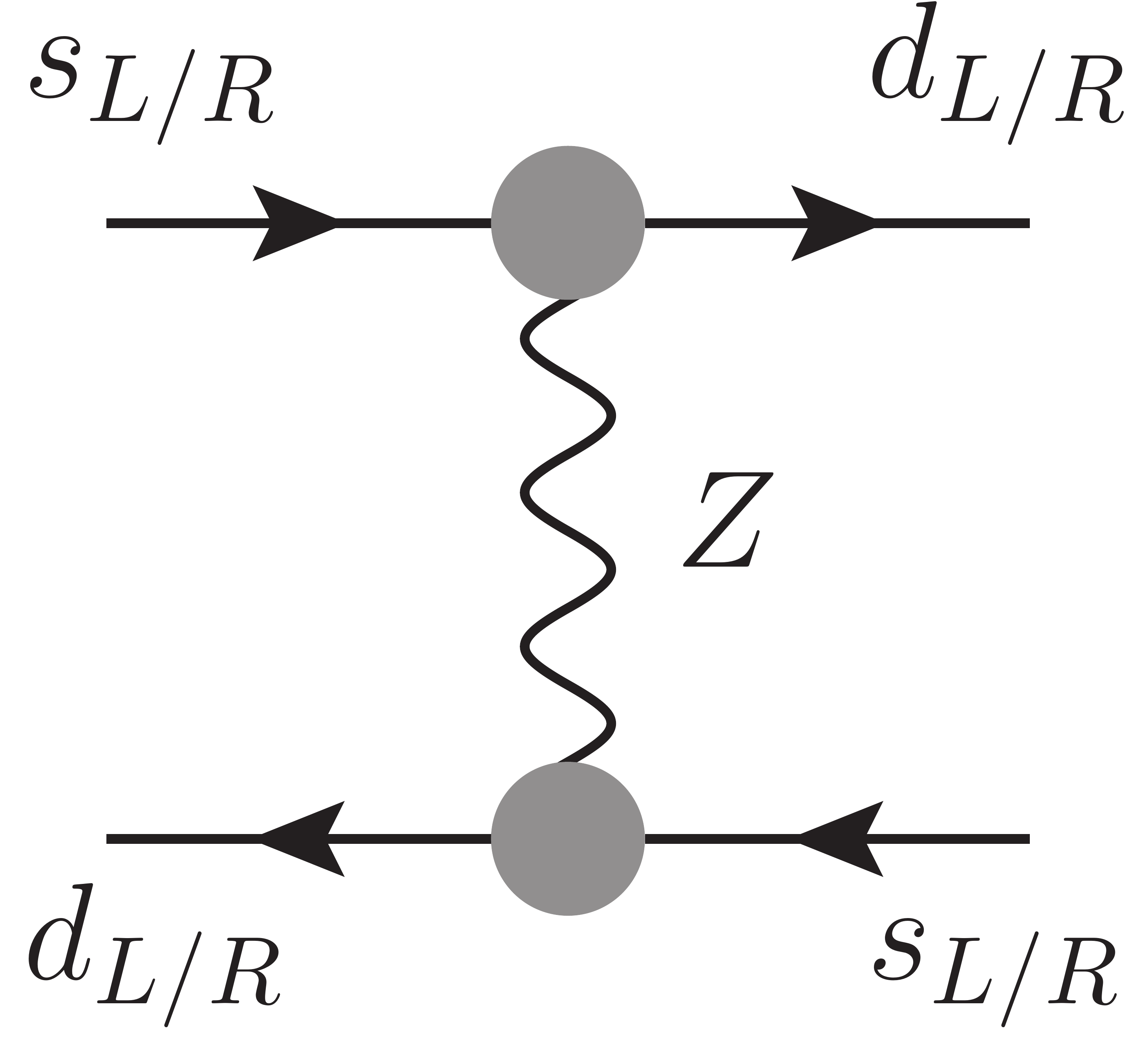}
}
\subfigure[]
{
\includegraphics[width=0.146\textwidth,bb = 0 0 831 748]{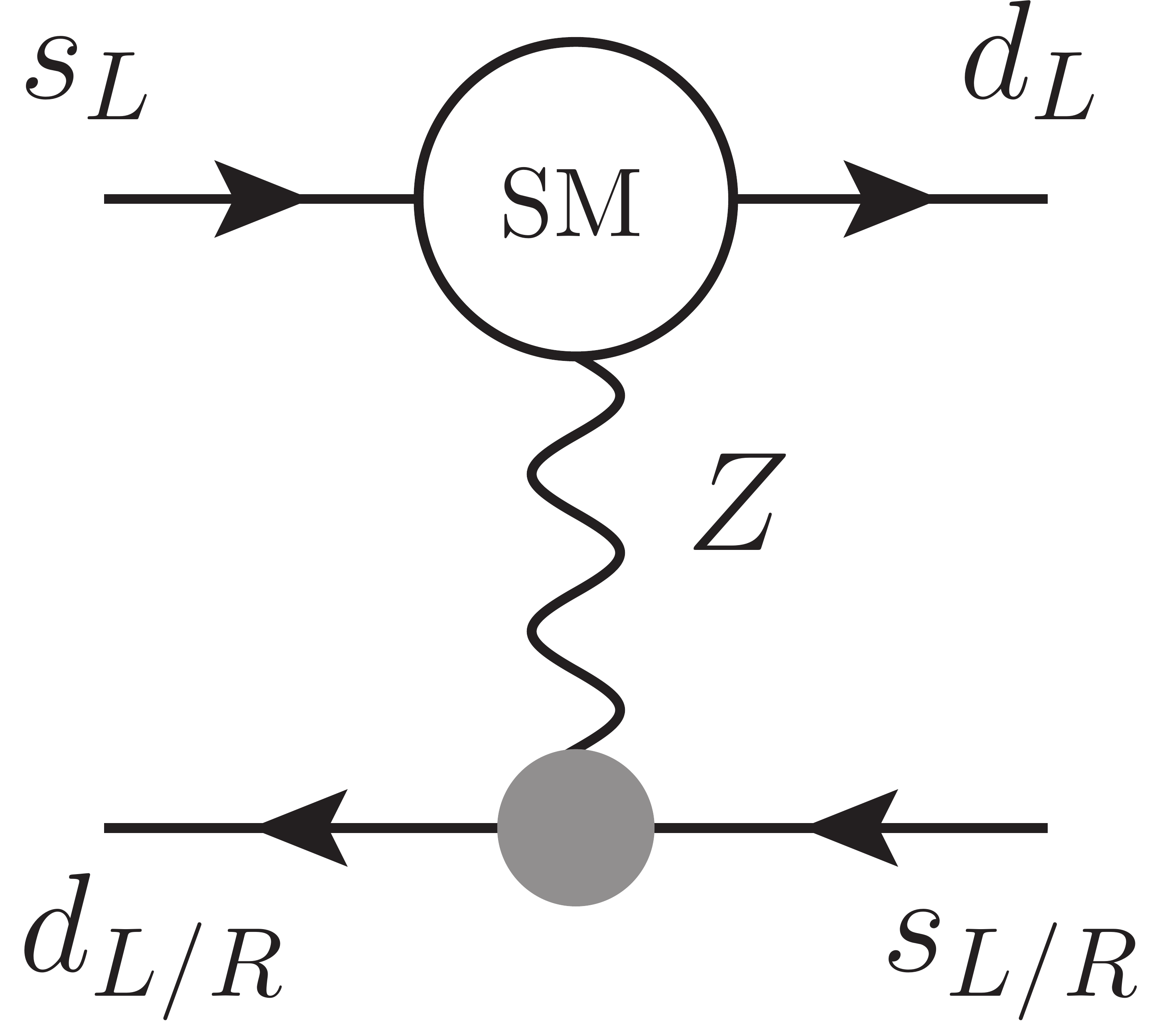}
}
\subfigure[]
{
\includegraphics[width=0.18\textwidth, bb = 0 0 1024 768]{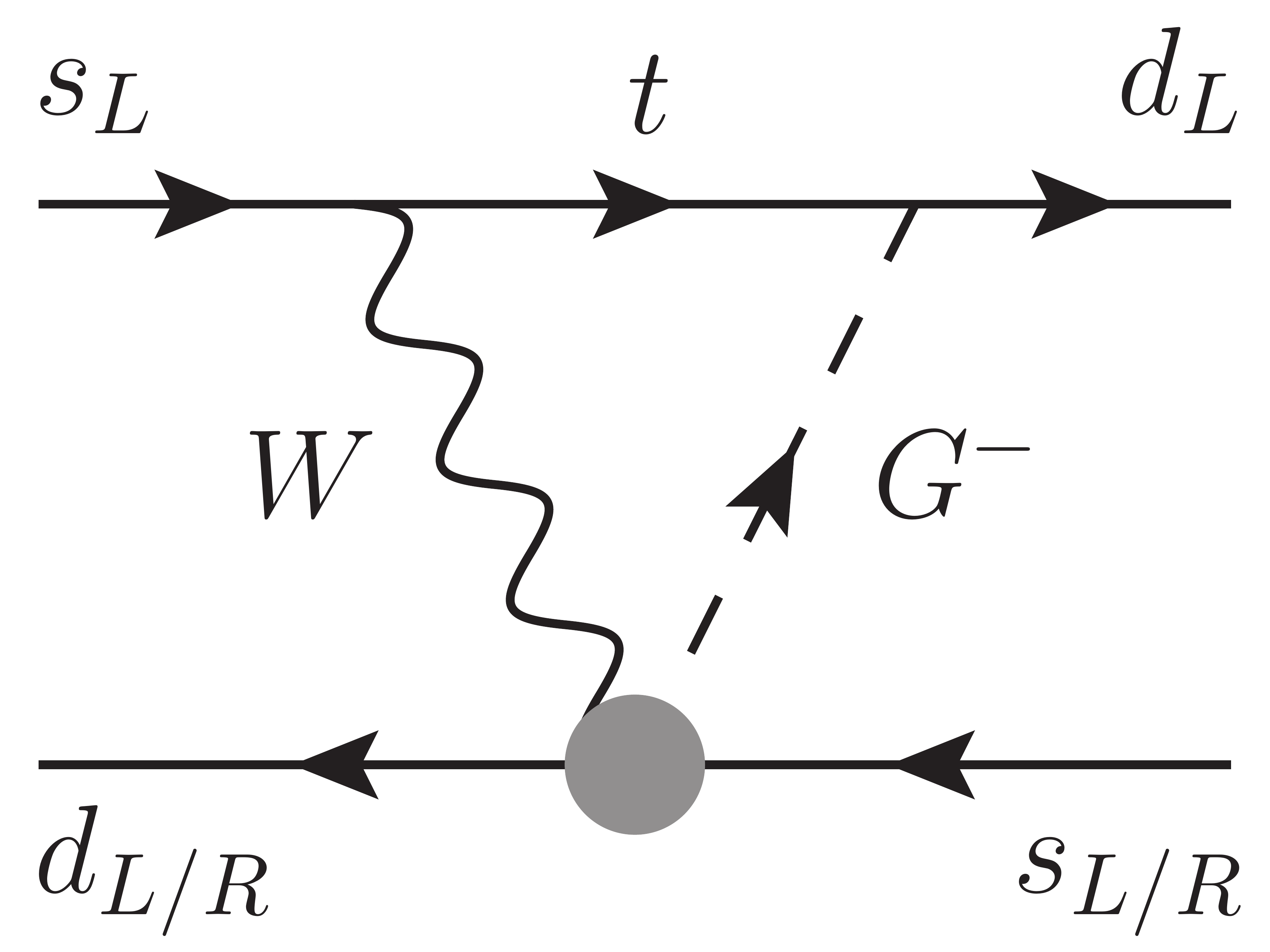}
}
\subfigure[]
{
\includegraphics[width=0.18\textwidth, bb = 0 0 1024 768]{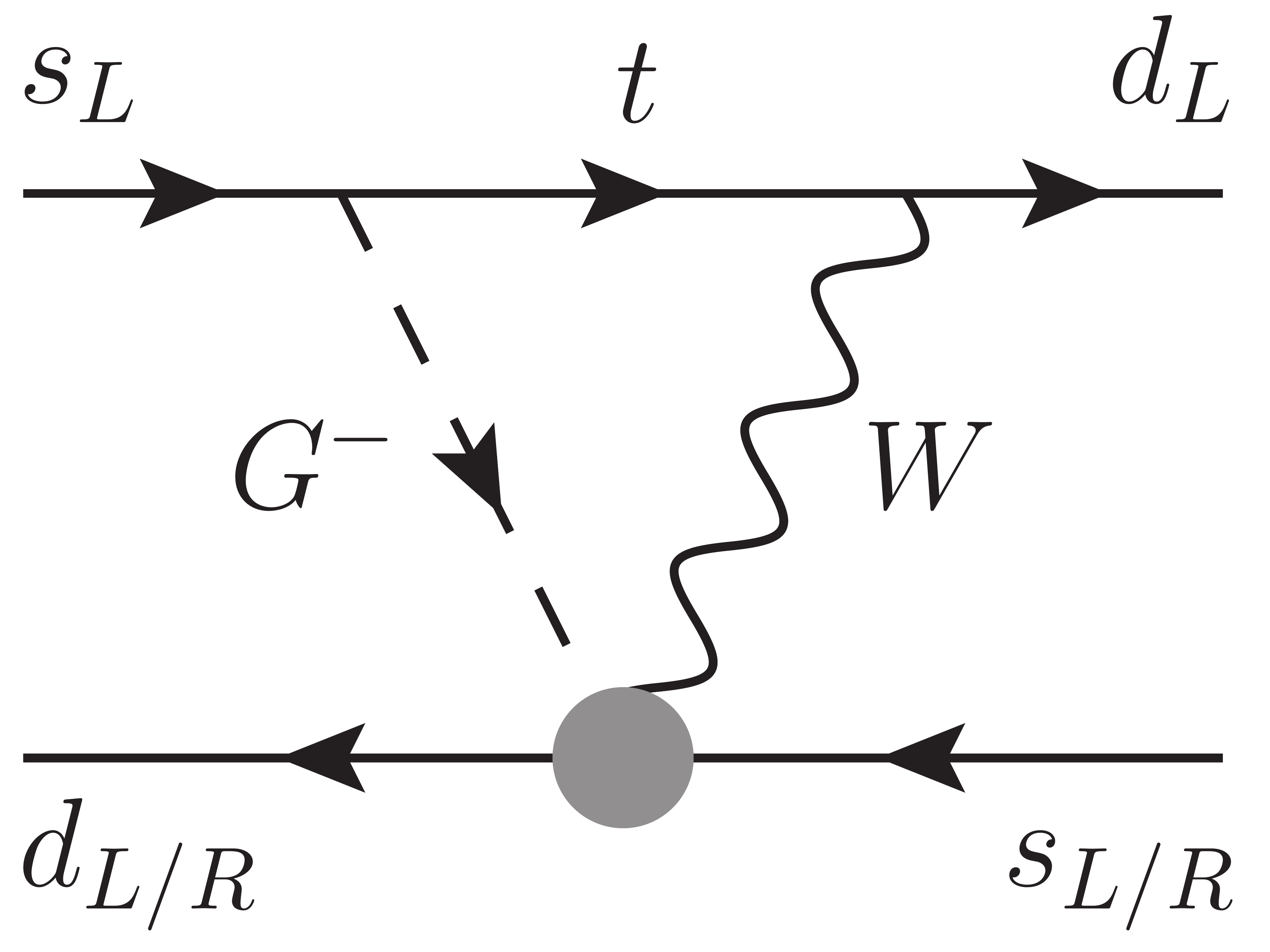}
}
\subfigure[]
{
\includegraphics[width=0.18\textwidth, bb = 0 0 1024 768]{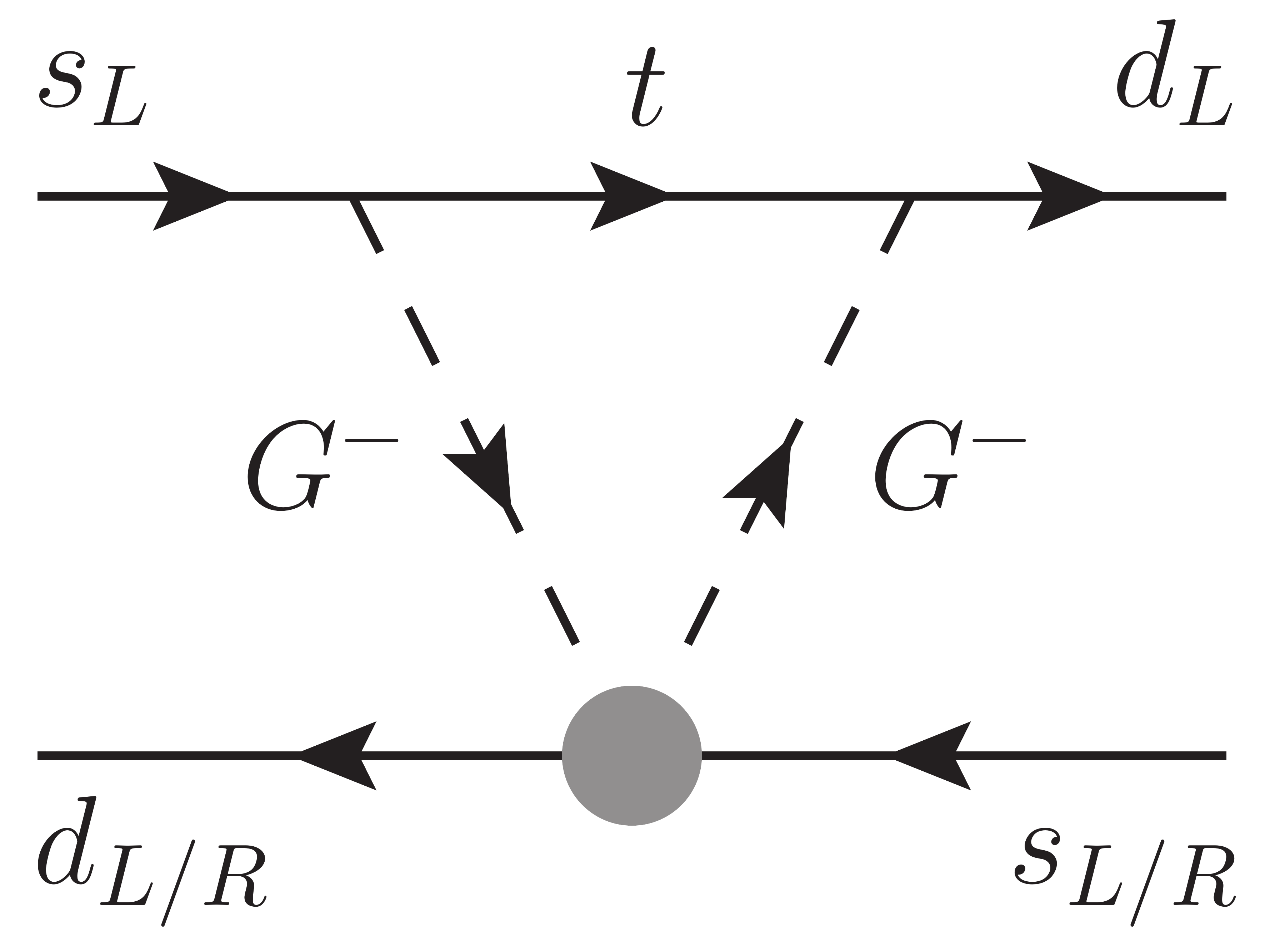}
}
\end{center}
\caption{
The NP contributions to $\Delta S = 2$ process. 
The black bubble denotes the vertices in Eq.~\eqref{eq:NPvertex} originating from the dimension-$6$ effective operators: $\mathcal{O}_L$ and $\mathcal{O}_R$.
The white bubble with ``SM" denotes the SM flavor-changing $Z$ interaction.
Subfigures~(b)--(e)  correspond to the interference contributions between the NP and SM.
A contribution from $G^0$-exchange diagram is negligible because it receives a suppression factor by the external momentum, so that we omit it here.
}
\label{fig:diagram}
\end{figure}

In the $\Delta S = 2$ observables, there are the indirect CP violation $\epsilon_K$ and the mass difference $\Delta m_K$ in the $\kk$ mixing. 
Since $\epsilon_K$ has been measured precisely, and the SM prediction is accurate, it provides a severe constraint.
The SM and NP contributions are shown as
\beq
 \epsilon_K &= e^{i \varphi_{\epsilon}}\left( \epsilon_K^{\rm SM} + \epsilon_K^{\rm NP} \right),
\eeq
where $\varphi_{\epsilon} =  (43.51 \pm 0.05)^{\circ}$.
The NP contribution is given by the double penguin diagrams with the $Z$ boson exchange (Fig.~\ref{fig:diagram}\,(a)),
\beq
  \epsilon_K^{\rm NP} &= \sum_{i=1}^{8}\left( \epsilon_K \right)^Z_i, 
\eeq
where the right-hand side is~\cite{Buras:2015jaq}
\begin{align}
 \left( \epsilon_K \right)^Z_1 &= -4.26\times10^7\, 
 {\rm Im}\,\Delta_L\,{\rm Re}\,\Delta_L,~~
 \left( \epsilon_K \right)^Z_2 = -4.26\times10^7\, 
 {\rm Im}\,\Delta_R\,{\rm Re}\,\Delta_R, \nonumber \\
 \left( \epsilon_K \right)^Z_3 &= 2.07\times10^9\, 
 {\rm Im}\,\Delta_L\,{\rm Re}\,\Delta_R,~~
 \left( \epsilon_K \right)^Z_4 = 2.07\times10^9\, 
 {\rm Im}\,\Delta_R\,{\rm Re}\,\Delta_L.
\end{align}
In these expressions, renormalization group corrections and long-distance contributions are included~\cite{Buras:2010pza}.
In addition, one must take account of the interference terms between the SM and NP contributions (Figs.~\ref{fig:diagram}\,(b)--(e)),
\begin{align}
 \left( \epsilon_K \right)^Z_5 &= -4.26\times10^7\, 
 {\rm Im}\,\Delta_L^{\rm SM}\,{\rm Re}\,\Delta_L,~~
 \left( \epsilon_K \right)^Z_6 = -4.26\times10^7\, 
 {\rm Im}\,\Delta_L\,{\rm Re}\,\Delta_L^{\rm SM}, \nonumber \\
 \left( \epsilon_K \right)^Z_7 &= 2.07\times10^9\, 
 {\rm Im}\,\Delta_L^{\rm SM}\,{\rm Re}\,\Delta_R,~~
 \left( \epsilon_K \right)^Z_8 = 2.07\times10^9\, 
 {\rm Im}\,\Delta_R\,{\rm Re}\,\Delta_L^{\rm SM}.
 \label{eq:ek_interference}
\end{align}
Here, the SM contribution, $\Delta_L^{\rm SM}$, is generated by radiative corrections.
%
%
%
%
At the one-loop level, it is calculated as
\beq
\Delta_L^{\rm SM} = \frac{g^3\lambda_t}{8 \pi^2 c_W}\widetilde C\!\left( \frac{m_t^2}{m_W^2},\mu_{\rm NP}  \right),
~~~
\Delta_R^{\rm SM} = 0,
\label{eq:DeltaLSM}
\eeq
where $c_W = \cos \theta_W$, $\lambda_i \equiv V_{is}^{\ast} V_{id}$ with the CKM matrix $V_{ij}$, and $\mu_{\rm NP}$ corresponds to the NP scale.\footnote{
In order to introduce how significant the interference contributions are, we ignore the renormalization group corrections to the dimension-6 operators above the electroweak scale except for a first leading logarithmic contribution $\ln (\mu_{\rm NP} /m_W) $ which comes from Fig.~\ref{fig:diagram}\,(e).
This approximation is valid when 
the NP scale is not so far from the electroweak scale.
}
In this letter, the \textsc{CKMfitter} result~\cite{Charles:2015gya} is used for the CKM elements, unless otherwise mentioned.
The loop function is\footnote{
  The loop function $\widetilde C(x,\mu_{\rm NP})$ is consistent with the result in Ref.~\cite{Aebischer:2015fzz}.
}
\beq
\widetilde C(x,\mu_{\rm NP}) = C(x) + \Delta C(x,\mu_{\rm NP}).
\eeq
In the Feynman-'t Hooft gauge, the first term in the right-hand side corresponds to the $Z$-boson exchange diagram~\cite{Buras:2012jb} (Fig.~\ref{fig:diagram}\,(b)),
\beq
C(x) = \frac{x}{8} \left[ \frac{x-6}{x-1} + \frac{3x+2}{(x-1)^2} \ln x \right],
\eeq
while the second term is obtained from the Nambu-Goldstone boson loops (Figs.~\ref{fig:diagram}\,(c)--(e)),
\beq
\Delta C(x ,\mu_{\rm NP}) = -\frac{x}{16} \left[ \frac{3x-17}{2(x-1)} - \frac{x(x-8)}{(x-1)^2} \ln x + \ln \frac{\mu_{\rm NP}^2}{m_W^2} \right].
\label{eq:NGboson}
\eeq
The explicit form of $\Delta C(x)$ depends on the effective operators above the electroweak symmetry breaking scale.
Here, they are supposed to be $\mathcal{O}_L$ and $\mathcal{O}_R$.
It is noted that the interference terms are gauge-independent.

The interference terms \eqref{eq:ek_interference} have been overlooked in Refs.~\cite{Buras:2012jb, Buras:2014sba, Buras:2015jaq, Buras:2015yca}.
They cannot be ignored, as we will see in the next section. 

The latest estimation of the SM value is~\cite{Ligeti:2016qpi}
\begin{align}
 \epsilon_K^{\rm SM} = (2.24 \pm 0.19) \times 10^{-3}.
\end{align}
On the other hand, the experimental result is~\cite{Olive:2016xmw}
\begin{align}
 |\epsilon_K^{\rm exp}| = (2.228 \pm 0.011) \times 10^{-3}.
\end{align}
They are well consistent with each other, and $\epsilon_K^{\rm NP}$ must satisfy
\begin{align}
 -0.39 \times 10^{-3} < \epsilon_K^{\rm NP} < 0.37 \times 10^{-3},
 \label{eq:ep_constraint}
\end{align}
at the $2\,\sigma$ level.\footnote{The SM estimation $\epsilon_K^{\rm SM}$ is sensitive to the CKM elements. 
If one uses $V_{cb}$ that is determined by the exclusive $B \to D^{(\ast)} \ell \nu$ decays~\cite{Bailey:2014tva},  $\epsilon_K^{\rm SM} = (1.73 \pm 0.18) \cdot 10^{-3} $ is obtained~\cite{Bailey:2015tba}. 
Then, $\epsilon_K^{\rm NP} = (0.50 \pm 0.18) \cdot 10^{-3}$ is required at the $1\,\sigma$ level.
}

The kaon mass difference $\Delta m_K$ consists of the SM and NP contributions:
\begin{align}
 \Delta m_K = \Delta m_K^{\rm SM} + \Delta m_K^{\rm NP}.
\end{align}
If we parameterize the NP contribution as
\beq
 \frac{\Delta m_K^{\rm NP}}{\Delta m_K^{\rm exp}} = \sum_{i=1}^8 R^Z_i,
\eeq
the right-hand side is estimated as~\cite{Buras:2015jaq}
\begin{align}
 R^Z_1 &= 6.43\times10^7\,\left[
 \left({\rm Re}\,\Delta_L\right)^2-
 \left({\rm Im}\,\Delta_L\right)^2 \right], \nonumber \\
 R^Z_2 &= 6.43\times10^7\,\left[
 \left({\rm Re}\,\Delta_R\right)^2-
 \left({\rm Im}\,\Delta_R\right)^2 \right], \nonumber \\
 R^Z_3 &= -6.21\times10^9\, {\rm Re}\,\Delta_L\,{\rm Re}\,\Delta_R,  \nonumber \\
 R^Z_4 &=  6.21\times10^9\, {\rm Im}\,\Delta_L\,{\rm Im}\,\Delta_R.
\end{align}
Similarly to the case of $\epsilon_K$, there are interference terms between the SM and NP contributions,
\begin{align}
 R^Z_5 &= 12.9\times10^7\, {\rm Re}\,\Delta_L^{\rm SM}\,{\rm Re}\,\Delta_L,~~
 R^Z_6 = -12.9\times10^7\, {\rm Im}\,\Delta_L^{\rm SM}\,{\rm Im}\,\Delta_L, \nonumber \\
 R^Z_7 &= -6.21\times10^9\, {\rm Re}\,\Delta_L^{\rm SM}\,{\rm Re}\,\Delta_R,~~
 R^Z_8 =  6.21\times10^9\, {\rm Im}\,\Delta_L^{\rm SM}\,{\rm Im}\,\Delta_R. 
 \label{eq:deltamk_interference}
\end{align}
Here, $\Delta_L^{\rm SM}$ is given by Eq.~\eqref{eq:DeltaLSM}, and the result is gauge-independent. 
These terms have been overlooked in the literature.

The experimental result is~\cite{Olive:2016xmw}
\begin{align}
\Delta m_K^{\rm exp} = (3.484 \pm 0.006) \times 10^{-15}\,{\rm GeV}.
\end{align}
Since the SM prediction involves sizable contributions of long-distance effects, the uncertainty is large.\footnote{
The latest lattice simulation, which includes the long-distance contributions, provides
$\Delta m_K^{\rm SM} = ( 3.19 \pm 1.04)\cdot 10^{-15}\,{\rm GeV}$~\cite{Bai:2014cva}.
However, it is performed on masses of unphysical pion, kaon and charmed quark.
} 
Hence, we simply require that the NP contribution does not exceed the experimental value (with the $2\,\sigma$ uncertainty):
\begin{align}
 |\Delta m_K^{\rm NP}| < 3.496 \times 10^{-15}\,{\rm GeV}.
\end{align}
This constraint will turn out to be much weaker than $\epsilon_K^{\rm NP}$.

\subsection{$\boldsymbol{\epsilon'/\epsilon}$}

The flavor-changing $Z$ interaction also contributes to $\Delta S = 1$ observables.
The direct CP violation $\epsilon'/\epsilon$ is shown as
\begin{align}
 \frac{\epsilon'}{\epsilon} = 
 \left( \frac{\epsilon'}{\epsilon} \right)_{\rm SM} +
 \left( \frac{\epsilon'}{\epsilon} \right)_{\rm NP}.
\end{align}
The NP contribution is estimated as~\cite{Buras:2015jaq}
\begin{align}
 \left( \frac{\epsilon'}{\epsilon} \right)_{\rm NP}
 = -2.64 \times 10^3\,B_8^{(3/2)} \left( {\rm Im}\,\Delta_L  
 + \frac{c_W^2}{s_W^2}\, {\rm Im}\,\Delta_R \right),
 \label{eq:epsprime}
\end{align}
where $B_8^{(3/2)} = 0.76 \pm 0.05$ from the lattice calculation.
Here, the terms which are not proportional to $B_8^{(3/2)}$ are omitted; 
the approximation is valid at the 10\,\% accuracy. 
A factor in the parenthesis gives $c_W^2/s_W^2 \simeq 3.33$.
Thus, the NP contribution can be enhanced easily by $\Delta_R$.

As mentioned in Sec.~\ref{sec:intro}, the SM prediction deviates from the experimental result at the $2.8$--$2.9\,\sigma$ level.
In this letter, we require that the discrepancy of $\epsilon'/\epsilon$ is explained at the $1\,\sigma$ level as
\begin{align}
 10.0\times 10^{-4} < \left(\frac{\epsilon'}{\epsilon}\right)_{\rm NP} < 21.1 \times 10^{-4},
 \label{eq:eppepNP}
\end{align}
where Ref.~\cite{Kitahara:2016nld} is used for the SM prediction.

\subsection{$\boldsymbol{K^+ \to \pi^+\nu\bar\nu}$ and $\boldsymbol{K_L \to \pi^0\nu\bar\nu}$}

The (ultra-)rare kaon decay channels, $K^+ \to \pi^+\nu\bar\nu$ and $K_L \to \pi^0\nu\bar\nu$, are correlated with $\epsilon'/\epsilon$ as well as $\epsilon_K$ and $\Delta m_K$ in the general $Z$ scenario.\footnote{
The branching ratios of $K\to \pi \ell^{+} \ell^{-}$ ($\ell = e,\,\mu$) are also affected in the general $Z$ scenario. 
However, $K^{+} \to \pi^{+} \ell^{+} \ell^{-} $ and $K_{S} \to \pi^{0} \ell^{+} \ell^{-} $ are dominated by a long-distance contribution through $K \to \pi \gamma^{\ast} \to  \pi \ell^{+} \ell^{-}$~\cite{Cirigliano:2011ny}. 
On the other hand, such a contribution to $K_L \to \pi^{0} \ell^{+} \ell^{-}$ is forbidden by the CP symmetry, but is dominated by an indirect CP-violating contribution, $K_L \to K_S \to \pi^{0} \ell^{+} \ell^{-}$~\cite{Cirigliano:2011ny}. 
Therefore, it is challenging to discuss shot-distance NP contributions in these channels.
}
They are represented as \cite{Buras:2015qea,Buras:2015jaq}
\begin{align}
 \mathcal{B}(K^+\to\pi^+\nu\bar\nu) &= 
 \kappa_+ \left[
 \left(
 \frac{{\rm Im}\,X_{\rm eff}}{\lambda^5} 
 \right)^2 + 
 \left(
 \frac{{\rm Re}\,\lambda_c}{\lambda}P_c(X) + 
 \frac{{\rm Re}\,X_{\rm eff}}{\lambda^5} 
 \right)^2
 \right], \\
 \mathcal{B}(K_L\to\pi^0\nu\bar\nu) &= 
 \kappa_L \left(
 \frac{{\rm Im}\,X_{\rm eff}}{\lambda^5} 
 \right)^2.
\end{align}
Here, $X_{\rm eff}$ is estimated as
\begin{align}
 X_{\rm eff} &=\lambda_t \left(1.48 \pm 0.01 \right)+
 2.51\times10^2\,\left(
 \Delta_L + \Delta_R
 \right), 
 \label{eq:Xeff}
\end{align}
where the first term in the right-hand side is the SM contribution. 
Also, $\lambda = |V_{us}|$, $\kappa_+ = (5.157 \pm 0.025) \cdot 10^{-11}(\lambda/0.225)^8$, and $\kappa_L = (2.231 \pm 0.013) \cdot 10^{-10}(\lambda/0.225)^8$.
The charm-quark contribution is $P_c(X)= (9.39 \pm 0.31)\cdot 10^{-4} /\lambda^4  + (0.04 \pm 0.02)$, where the first term in the right-hand side comes from short-distance effects, while the second one takes account of long-distance effects. 
Using the \textsc{CKMfitter} result for the CKM elements, one obtains
\begin{align}
 {\rm Re}\,X_{\rm eff} &= -4.83\times10^{-4}+
 2.51\times10^2\,\left(
 {\rm Re}\,\Delta_L + {\rm Re}\,\Delta_R
 \right), 
 \label{eq:SMkplus}\\
 {\rm Im}\,X_{\rm eff} &= 2.12\times10^{-4}+
 2.51\times10^2\,\left(
 {\rm Im}\,\Delta_L + {\rm Im}\,\Delta_R
 \right).
 \label{eq:SML}
\end{align}
The SM predictions become 
\begin{align}
 \mathcal{B}(K^+\to\pi^+\nu\bar\nu)_{\rm SM} &= (8.5 \pm 0.5) \times 10^{-11},  \label{eq:KppinunuSM} \\
 \mathcal{B}(K_L\to\pi^0\nu\bar\nu)_{\rm SM} &= (3.0 \pm 0.2) \times 10^{-11}.
 \label{eq:KpinunuSM}
\end{align}
On the other hand, the experimental results are~\cite{Artamonov:2008qb, Ahn:2009gb}
\begin{align}
 \mathcal{B}(K^+\to\pi^+\nu\bar\nu)_{\rm exp} &= (17.3^{+11.5}_{-10.5}) \times 10^{-11}, 
 \label{eq:Kpexp} \\
 \mathcal{B}(K_L\to\pi^0\nu\bar\nu)_{\rm exp} & \leq 2.6 \times 10^{-8}.~~~[90\%~\mbox{C.L.}]
\end{align}
Although the current constraints on the NP contributions are very weak, the experimental values will be improved significantly in the near future. 
The NA62 experiment at CERN, which already started the physics run at low beam intensity in 2015, has a potential to measure $\mathcal{B}(K^+\to\pi^+\nu\bar\nu)$ at the 10\,\% precision by 2018~\cite{Moulson:2016dul}.
The KOTO experiment at J-PARC is designed to improve the sensitivity for $\mathcal{B}(K_L\to\pi^0\nu\bar\nu)$, which enables us to measure it at the 10\,\% level of the SM value~\cite{Tung:2016xtx,KOTO}.
As one can see from Eqs.~\eqref{eq:epsprime} and \eqref{eq:Xeff}, the NP contributions to $\mathcal{B}(K \to \pi \nu \bar{\nu})$ are correlated with those to $\epsilon'/\epsilon$ in the general $Z$ scenario.
Thus, if the $\epsilon'/\epsilon$ discrepancy is a signal of the scenario, these experiments would detect  NP effects.

\subsection{$\boldsymbol{K_L \to \mu^+\mu^-}$}

The branching ratio of $K_L \to \mu^+\mu^-$ is also sensitive to the NP contributions to the flavor-changing $Z$ couplings. 
Theoretically, only the short-distance (SD) contributions can be calculated reliably.
They are shown as~\cite{Gorbahn:2006bm,Bobeth:2013tba,Buras:2015jaq}
\begin{align}
 \mathcal{B}(K_L \to \mu^+\mu^-)_{\rm SD} &= 
 \kappa_\mu 
 \left(
 \frac{{\rm Re}\,\lambda_c}{\lambda}P_c(Y) + 
 \frac{{\rm Re}\,Y_{\rm eff}}{\lambda^5} 
 \right)^2,
\end{align}
where $\kappa_\mu=(2.01\pm0.02)\cdot10^{-9}(\lambda/0.225)^8$. 
The charm-quark contribution is $P_c(Y)= (0.115 \pm 0.018)\cdot (0.225/\lambda)^4$. 
Using the \textsc{CKMfitter} result, one obtains
\begin{align}
 {\rm Re}\,Y_{\rm eff} &= -3.07\times10^{-4}+
 2.51\times10^2\,\left(
 {\rm Re}\,\Delta_L - {\rm Re}\,\Delta_R
 \right),
\end{align}
where the first term in the right-hand side is the SM contribution, and the minus sign between $\Delta_L$ and $\Delta_R$ is due to the axial-vector current.
The SM value is obtained as
\begin{align}
 \mathcal{B}(K_L \to \mu^+\mu^-)_{\rm SD,\,SM} = (0.83 \pm 0.10) \times 10^{-9}.
\end{align}
On the other hand, it is challenging to extract a short-distance part in the experimental data $\mathcal{B}(K_L \to \mu^{+} \mu^{-})_{\rm exp} = (6.84 \pm 0.11)\cdot 10^{-9}$~\cite{Olive:2016xmw}, because of huge long-distance contributions through $K_L \to \gamma^{\ast} \gamma^{\ast} \to \mu^{+} \mu^{-}$~\cite{Isidori:2003ts}.
An upper bound on the short-distance contribution is~\cite{Isidori:2003ts}
\begin{align}
 \mathcal{B}(K_L \to \mu^+\mu^-)_{\rm SD} < 2.5 \times 10^{-9}.
\end{align}
Since the constraint is much weaker than the SM uncertainties, we ignore them for simplicity and impose a bound on the $Z$ couplings,
\beq
- 1.08 \times 10^{-6} <  {\rm Re}\,\Delta_L - {\rm Re}\,\Delta_R < 4.05 \times  10^{-6}.
\eeq
The real parts of the NP contributions are constrained by $\mathcal{B} (K_L \to \mu^{+} \mu^{-} )$.

\section{Analysis}
\setcounter{equation}{0}

In this section, we examine the general $Z$ scenario. 
Although the discrepancy of $\epsilon'/\epsilon$ could be explained by the scenario, the parameter regions would be constrained by $\epsilon_K$, $\Delta m_K$ and $K_L \to \mu^+\mu^-$.
In particular, the interference between the SM and NP contributions, Eq.~\eqref{eq:ek_interference}, affects $\epsilon_K$ significantly.
In this section, we choose the NP scale, $\mu_{\rm NP} =1\TeV$, as a reference.
As we will see, wide parameter regions are excluded.
Thus, the discrepancy of $\epsilon'/\epsilon$ will be explained by tuning the model parameters.
Let us introduce a quantity which parameterizes the tuning:\footnote{
  Our definition is almost the same as that in Ref.~\cite{Blanke:2008zb}, where the authors discuss correlations between the tuning parameter and flavor observables. 
}
\begin{align}
\xi = \textrm{max}\bigl( \xi_1, \xi_2, \dots, \xi_8 \bigr), \textrm{~~~~~with~~}
\xi_i = \left|\frac{\left( \epsilon_K \right)^Z_i}{\epsilon_K^{\rm NP}}\right|.
\label{eq:tuning}
\end{align}
If $\epsilon_K^{\rm NP}$ is dominated by a single term, one obtains $\xi \simeq 1$ and there is no tuning among the model parameters. 
If the maximal value of $\left( \epsilon_K \right)^Z_i$ is about ten times larger than $\epsilon_K^{\rm NP}$, $\xi \sim 10$ is obtained; the model parameters are tuned such that there is a cancellation among $\left( \epsilon_K \right)^Z_i$ at the 10\% level.

\subsection{Simplified scenarios}

\begin{figure}[t]
\begin{center}
\subfigure[Left-handed scenario (LHS)]
{
\includegraphics[width=0.47\textwidth, bb = 0 0 338 322]{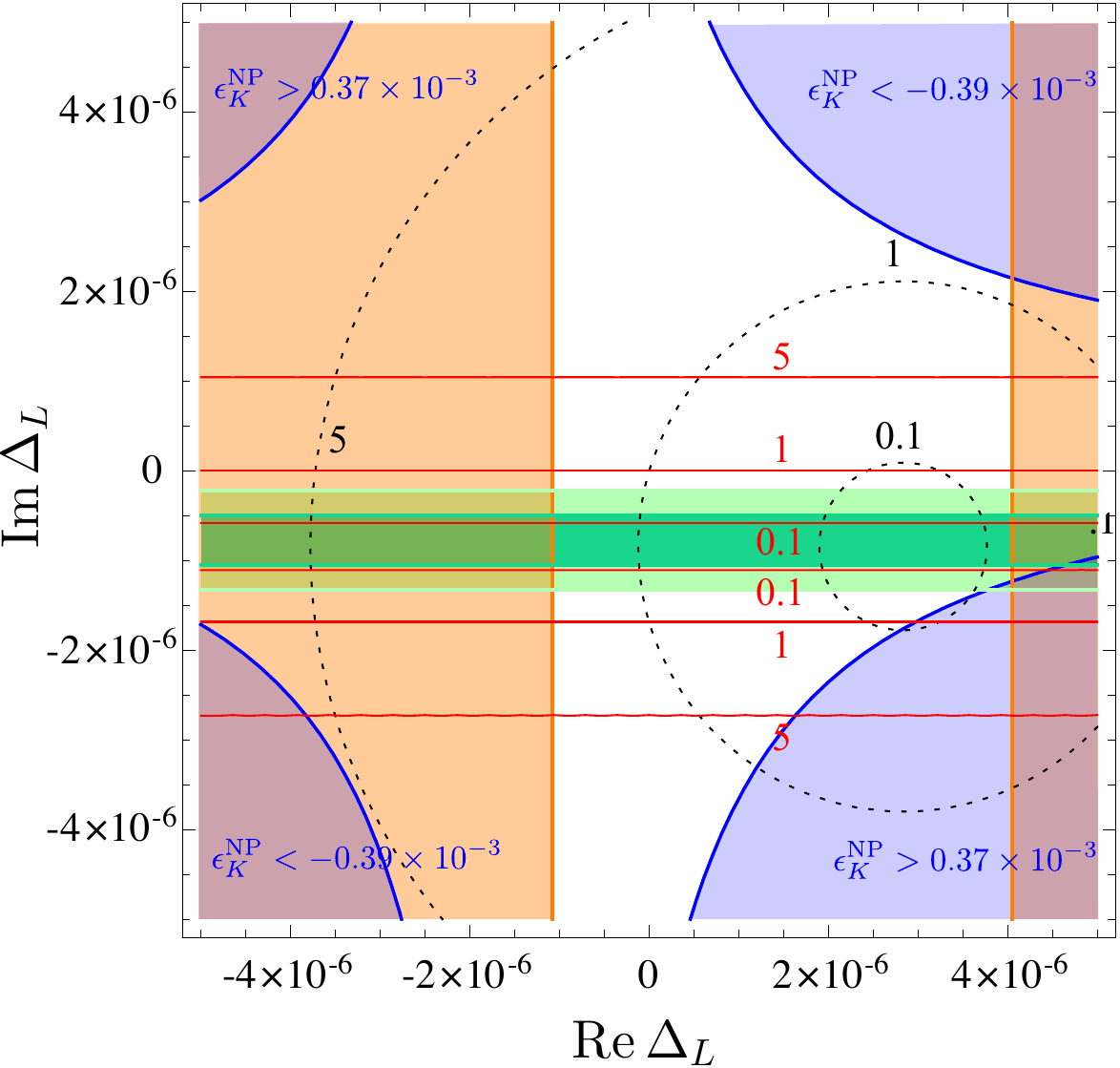}
}
~
\subfigure[Right-handed scenario (RHS)]
{
\includegraphics[width=0.47\textwidth, bb = 0 0 338 322]{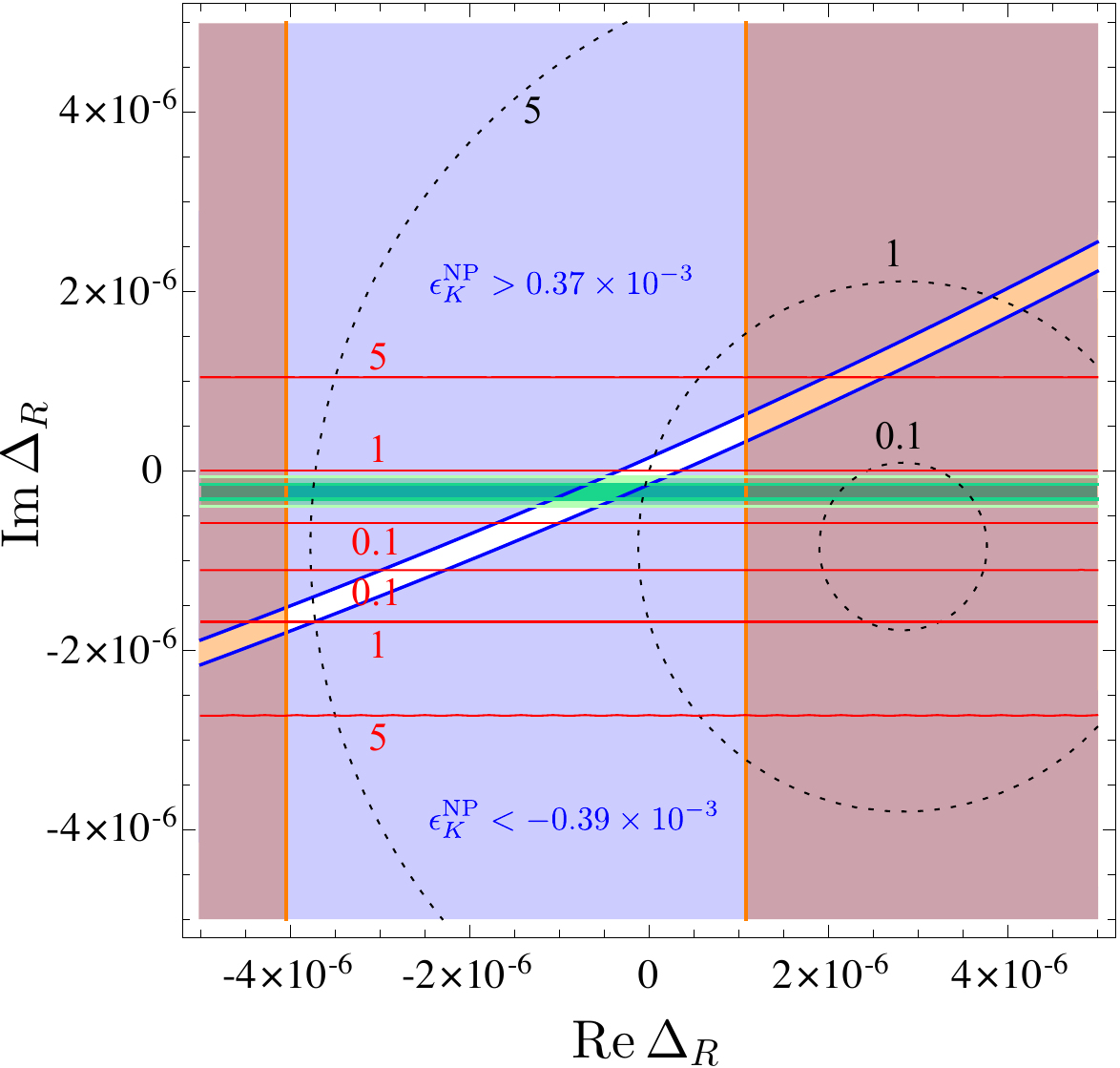}
}
\end{center}
\caption{
The $Z$-penguin observables are displayed in LHS ({\em left} panel) and RHS ({\em right}). 
In the green (light green) regions, the $\epsilon' / \epsilon$ discrepancy is explained at $1\,(2)\,\sigma$.
The blue and the orange shaded regions are excluded by $\epsilon_K$ and $ \mathcal{B}(K_L \to \mu^{+} \mu^{-})$, respectively.
The ratios of $\mathcal{B}(K_L \to \pi^{0} \nu \bar{\nu})/\mathcal{B}(K_L \to \pi^{0} \nu \bar{\nu})_{\rm SM}$ and $\mathcal{B}(K^{+} \to \pi^{+} \nu \bar{\nu})/\mathcal{B}(K^{+} \to \pi^{+} \nu \bar{\nu})_{\rm SM}$ are shown by the red solid and black dashed contours, respectively.
The NP scale is $\mu_{\rm NP} =1\TeV$.
}
\label{fig:LHS_RHS}
\end{figure}

First, we consider the following simplified scenarios (c.f., Ref.~\cite{Buras:1998ed}),
\begin{itemize}
\item left-handed scenario (LHS): $\Delta_R = 0$,\footnote{
 This scenario is realized by chargino contributions to the $Z$ penguin in the supersymmetric model~\cite{Buras:1997ij,Colangelo:1998pm,Buras:1999da,Tanimoto:2016yfy,Endo:2016aws}.
}
\item right-handed scenario (RHS): $\Delta_L = 0$,\footnote{
 Randall-Sundrum models with custodial protection~\cite{Blanke:2008zb,Blanke:2008yr} can generate large $\Delta_R$. 
 However, there are additional effects, e.g., from KK-gluon diagrams for $\epsilon_K^{\rm NP}$.
}
\item pure imaginary scenario (ImZS): ${\rm Re}\,\Delta_L = {\rm Re}\,\Delta_R = 0$,
\item left-right symmetric scenario (LRS): $\Delta_L = \Delta_R$.\footnote{
In axial-symmetric scenarios, $\Delta_L = -\Delta_R$, there are no NP contributions to $K\to\pi\nu\bar\nu$. 
}
\end{itemize}
As shown below, these scenarios do not require large parameter tuning in $\epsilon_K^{\rm NP}$.
However, $\mathcal{B}(K \to \pi \nu \bar{\nu})$ will turn out to be small.

In Fig.~\ref{fig:LHS_RHS}, the $Z$-penguin observables are shown as functions of $\Delta_{L,R}$ for LHS and RHS.
In the green (light green) regions, the $\epsilon' / \epsilon$ discrepancy is explained at $1\,(2)\,\sigma$.
They depend only on the imaginary component of $\Delta_{L,R}$. 
Obviously, $\epsilon' / \epsilon$ is enhanced by the right-handed $Z$ coupling, $\Delta_{R}$, more than $\Delta_{L}$.

The blue regions are excluded by the $\epsilon_K$, and the orange regions are by the $\mathcal{B}(K_L \to \mu^{+} \mu^{-})$.
The constraint from $\epsilon_K$ is much severer in RHS than LHS due to the interference contributions, Eq.~\eqref{eq:ek_interference}.
There is no constraint from $\Delta m_K$ in the parameter regions of the plots. 

The red and black dashed contours represent $\mathcal{B}(K_L \to \pi^{0} \nu \bar{\nu})/\mathcal{B}(K_L \to \pi^{0} \nu \bar{\nu})_{\rm SM}$ and $\mathcal{B}(K^{+} \to \pi^{+} \nu \bar{\nu})/\mathcal{B}(K^{+} \to \pi^{+} \nu \bar{\nu})_{\rm SM}$, respectively.
Here and hereafter, $\mathcal{B}(K_L \to \pi^{0} \nu \bar{\nu})_{\rm SM}$ and $\mathcal{B}(K^{+} \to \pi^{+} \nu \bar{\nu})_{\rm SM}$ denote the central values of the SM predictions, Eqs.~\eqref{eq:KppinunuSM} and \eqref{eq:KpinunuSM}.
It is found that $\mathcal{B}(K_L \to \pi^{0} \nu \bar{\nu})$ cannot be as large as the SM value as long as $\epsilon' / \epsilon$ is explained in LHS or RHS.
On the other hand, if the $\epsilon' / \epsilon$ discrepancy is explained by LHS, the NP contribution to $\mathcal{B}(K^{+} \to \pi^{+} \nu \bar{\nu})$ is limited by $\mathcal{B}(K_L \to \mu^{+} \mu^{-})$. 
In contrast, $\epsilon_K$ restricts RHS.

\begin{figure}[tp]
\begin{center}
\subfigure[Pure imaginary scenario (ImZS)]
{
\includegraphics[width=0.47\textwidth, bb = 0 0 338 322]{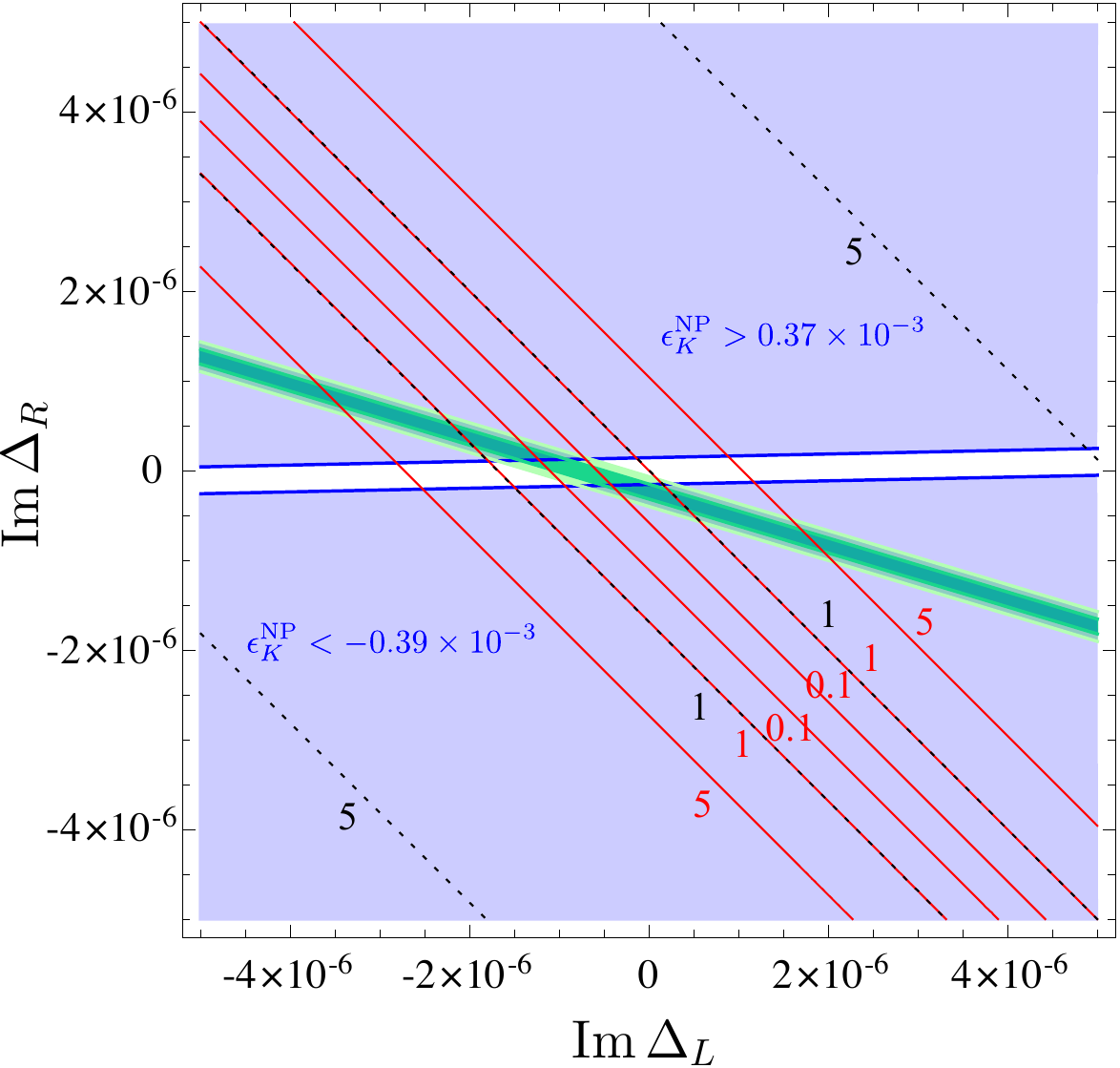}
}
~
\subfigure[Left-right symmetric scenario (LRS)]
{
\includegraphics[width=0.47\textwidth, bb = 0 0 338 322]{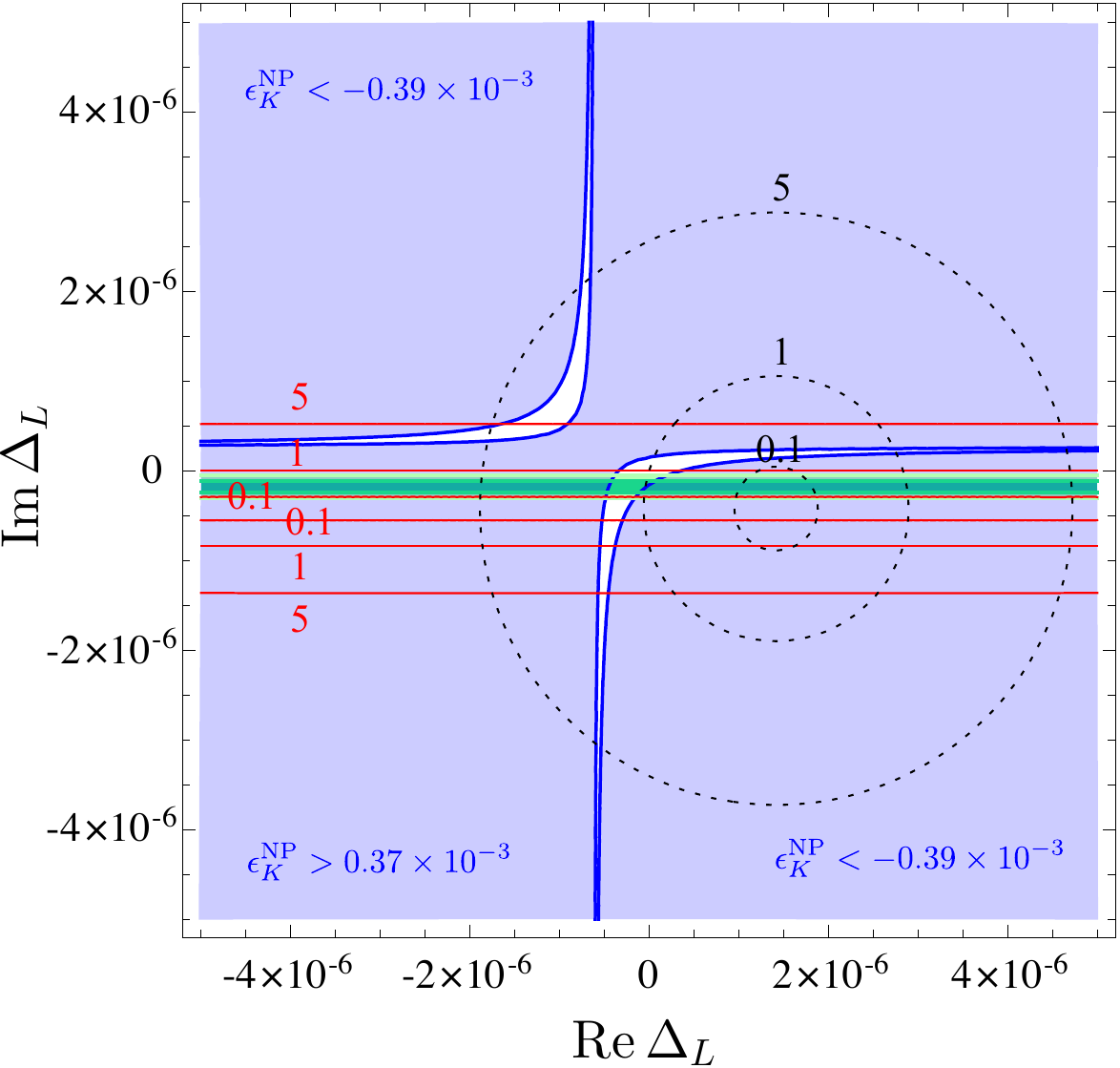}
}
\end{center}
\caption{
The $Z$-penguin observables are displayed in ImZS ({\em left} panel) and LRS ({\em right}). 
Notations of the lines and shaded regions are the same as in Fig.~\ref{fig:LHS_RHS}.
}
\label{fig:ImZS}
\end{figure}

Next, we consider ImZS.
Such a situation is often considered to amplify $(\epsilon' /\epsilon)_{\rm NP}$ but suppress $\epsilon_K^{\rm NP}$.
In the left panel of Fig.~\ref{fig:ImZS}, the $Z$-penguin observables are shown as functions of ${\rm Im}\,\Delta_{L,R}$.
The most severe constraint is from $\epsilon_K$ due to the interference between the SM and NP.
The other bounds are weak and absent in the plot.  
Since there are no real components of $\Delta_{L,R}$, $\mathcal{B}(K^{+} \to \pi^{+} \nu \bar{\nu})$ is correlated with $\mathcal{B}(K_L \to \pi^{0} \nu \bar{\nu})$.

Finally, LRS is shown in the right panel of Fig.~\ref{fig:ImZS}.
Similarly to the cases of RHS and ImZS, most of the parameter regions are excluded by $\epsilon_K$. 
The NP contributions to $\mathcal{B}(K_L \to \mu^{+} \mu^{-})$ vanish because the process is the axial-vector current.

\begin{figure}[t]
\begin{center}
\subfigure[LHS, RHS, and ImZS]
{
\includegraphics[width=0.47\textwidth, bb = 0 0 561 472]{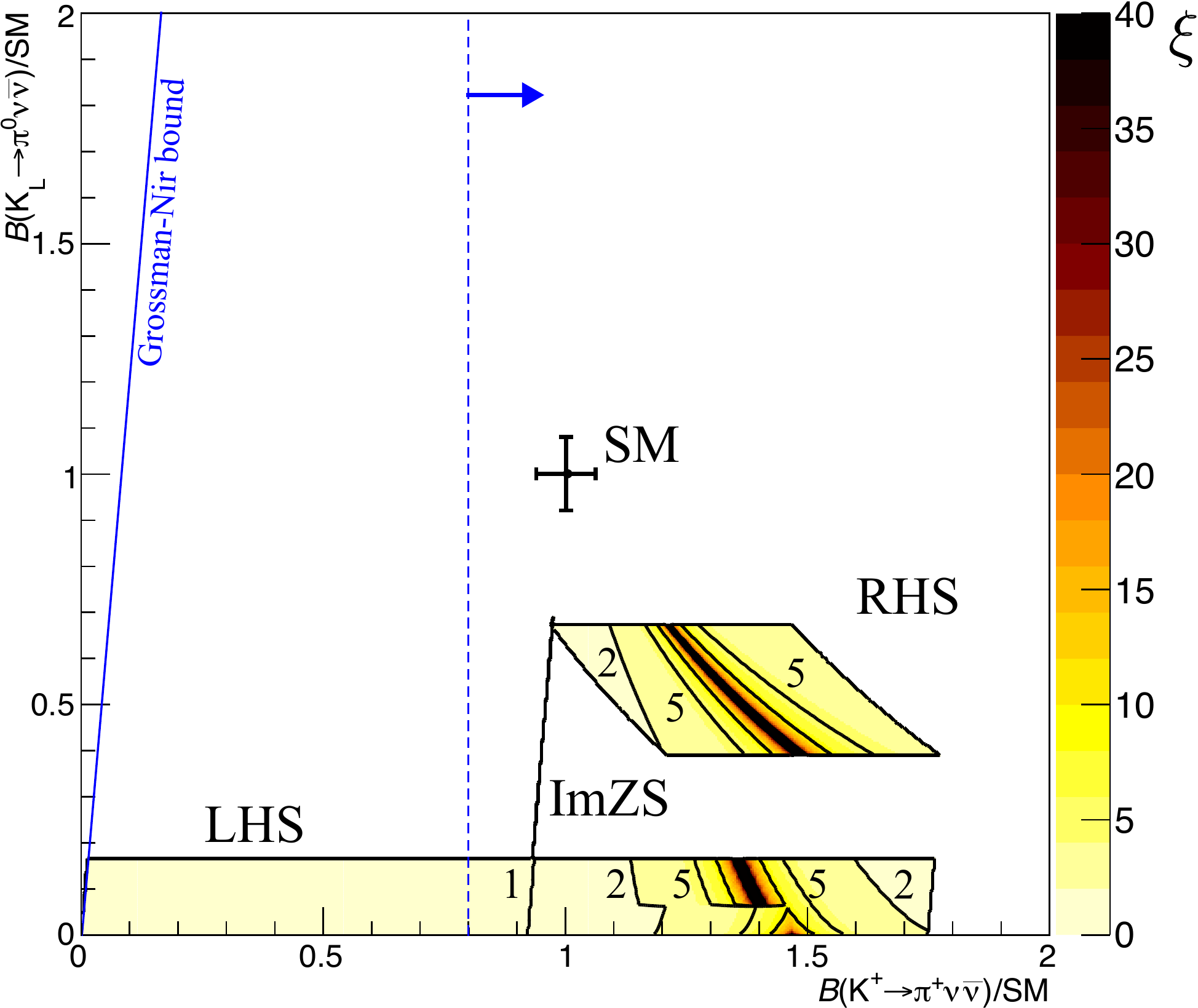}
}
\subfigure[LRS]
{
\includegraphics[width=0.47\textwidth, bb = 0 0 558 472]{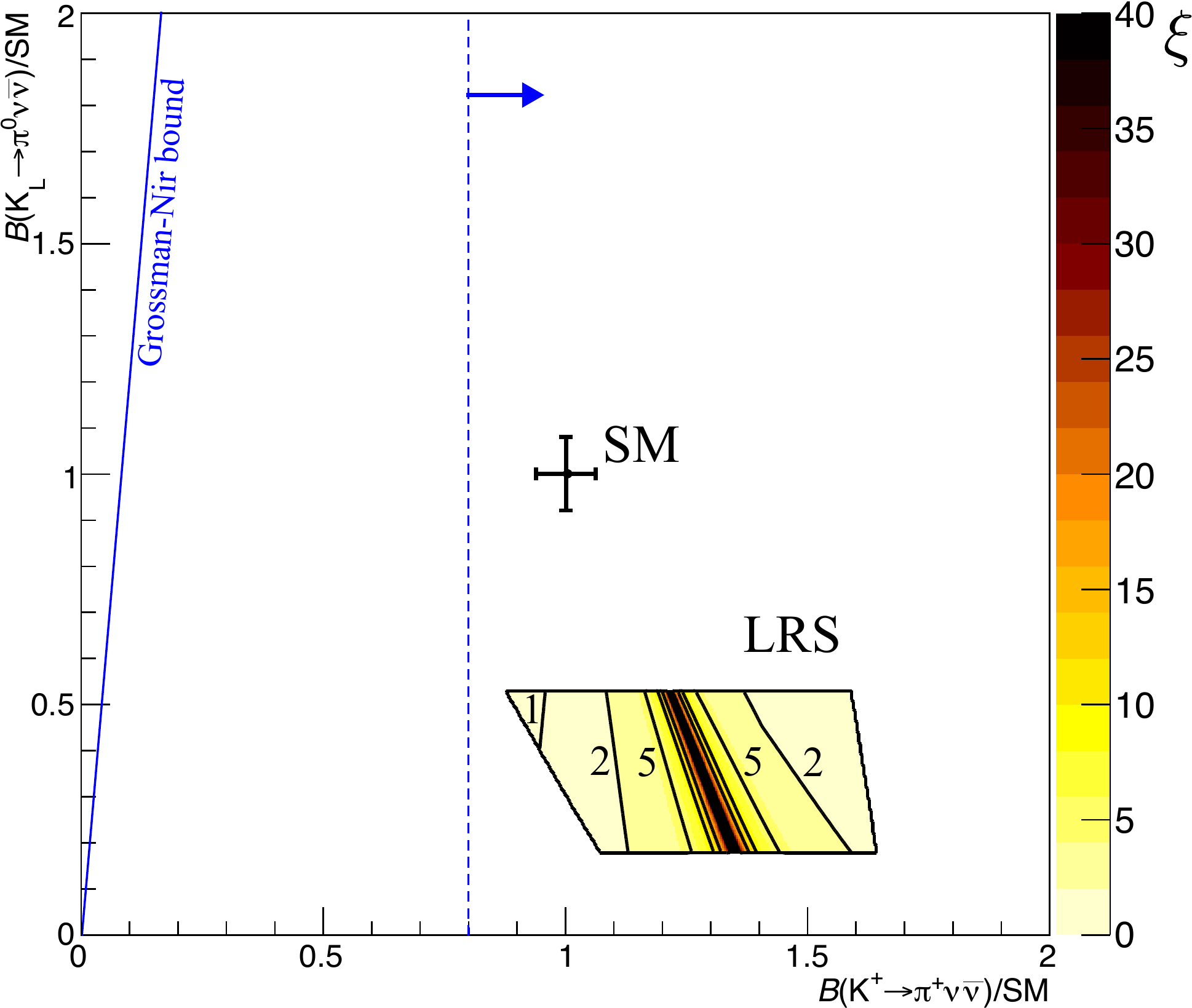}
}
\end{center}
\caption{
Contours of the tuning parameter $\xi$ are shown in the simplified scenarios: LHS, RHS, and ImZS ({\em left} panel) and LRS ({\em right}).
Here, ``SM'' in the axis labels denotes the central values of $\mathcal{B}(K_L \to \pi^{0} \nu \bar{\nu})_{\rm SM}$ and $\mathcal{B}(K^{+} \to \pi^{+} \nu \bar{\nu})_{\rm SM}$, Eqs.~\eqref{eq:KppinunuSM} and \eqref{eq:KpinunuSM}.
In the colored regions, $\epsilon' / \epsilon$ is explained at $1\sigma$, and the experimental bounds of $\epsilon_K$, $\Delta m_K$, and $\mathcal{B}(K_L \to \mu^{+} \mu^{-})$ are satisfied.
The right region of the blue dashed line is allowed by the measurement of $\mathcal{B}(K^{+} \to \pi^{+} \nu \bar{\nu})$ at $1\sigma$.
The Grossman-Nir bound~\cite{Grossman:1997sk} is shown by the blue solid line.
The NP scale is set to be $\mu_{\rm NP} =1$\,TeV.
}
\label{fig:simplified}
\end{figure}

In Fig.~\ref{fig:simplified}, contours of the tuning parameter $\xi$ are shown for the simplified scenarios:  LHS, RHS, ImZS, and LRS on the plane of the branching ratios of $K \to \pi \nu \bar{\nu}$.
We scanned the whole parameter space of $\Delta_{L,R}$ in each scenario and selected the parameters where $\epsilon' / \epsilon$ is explained at the $1\sigma$ level, and the experimental bounds from $\epsilon_K$, $\Delta m_K$, and $\mathcal{B}(K_L \to \mu^{+} \mu^{-})$ are satisfied (see the previous section for the experimental constraints). 
Then, $\xi$ was estimated at each point.
Several parameter sets predict the same $\mathcal{B}(K^{+}\to \pi^{+}\nu \bar{\nu})$ and $\mathcal{B}(K_{L}\to \pi^{0} \nu \bar{\nu} )$.
Among them, the smallest $\xi$ is chosen in Fig.~\ref{fig:simplified} for each set of $\mathcal{B}(K^{+}\to \pi^{+}\nu \bar{\nu})$ and $\mathcal{B}(K_{L}\to \pi^{0} \nu \bar{\nu} )$. 
In most of the allowed parameter regions, $\xi = \mathcal{O}(1)$ is obtained.
Thus, one does not require tight tunings in these scenarios. 

In the figures, $\mathcal{B}(K_{L}\to \pi^{0} \nu \bar{\nu} )$ is smaller than the SM value by more than 30\%. 
Hence, the scenarios could be tested by the KOTO experiment.
On the other hand, $\mathcal{B}(K^{+}\to \pi^{+}\nu \bar{\nu})$ depends on the scenarios. 
In LHS, we obtain $0 < \mathcal{B}(K^{+}\to \pi^{+}\nu \bar{\nu})/ \mathcal{B}(K^{+}\to \pi^{+}\nu \bar{\nu})_{\rm SM} < 1.8$. 
In RHS, $\mathcal{B}(K^{+}\to \pi^{+}\nu \bar{\nu})$ is comparable to or larger than the SM value, but cannot be twice as large.
In ImZS, the branching ratios are perfectly correlated and displayed by a line in Fig.~\ref{fig:simplified}. 
Then, $\mathcal{B}(K^{+}\to \pi^{+}\nu \bar{\nu})$ is not deviated from the SM one.  
The right panel of Fig.~\ref{fig:simplified} is a result of the tuning parameter $\xi$ in LRS.
It is found that $\mathcal{B}(K_{L}\to \pi^{0} \nu \bar{\nu} )$ does not exceed about a half of the SM value.
On the other hand, $\mathcal{B}(K^{+}\to \pi^{+}\nu \bar{\nu})$ is comparable to or larger than the SM value, but cannot be twice as large, as is similar to RHS.

\subsection{General scenario}

\begin{figure}[tp]
\begin{center}
\subfigure[$\mathcal{B}(K^{+} \to \pi^{+} \nu \bar{\nu})/\mathcal{B}(K^{+} \to \pi^{+} \nu \bar{\nu})_{\rm SM}$]
{
\includegraphics[width=0.465\textwidth, bb = 0 0 338 322]{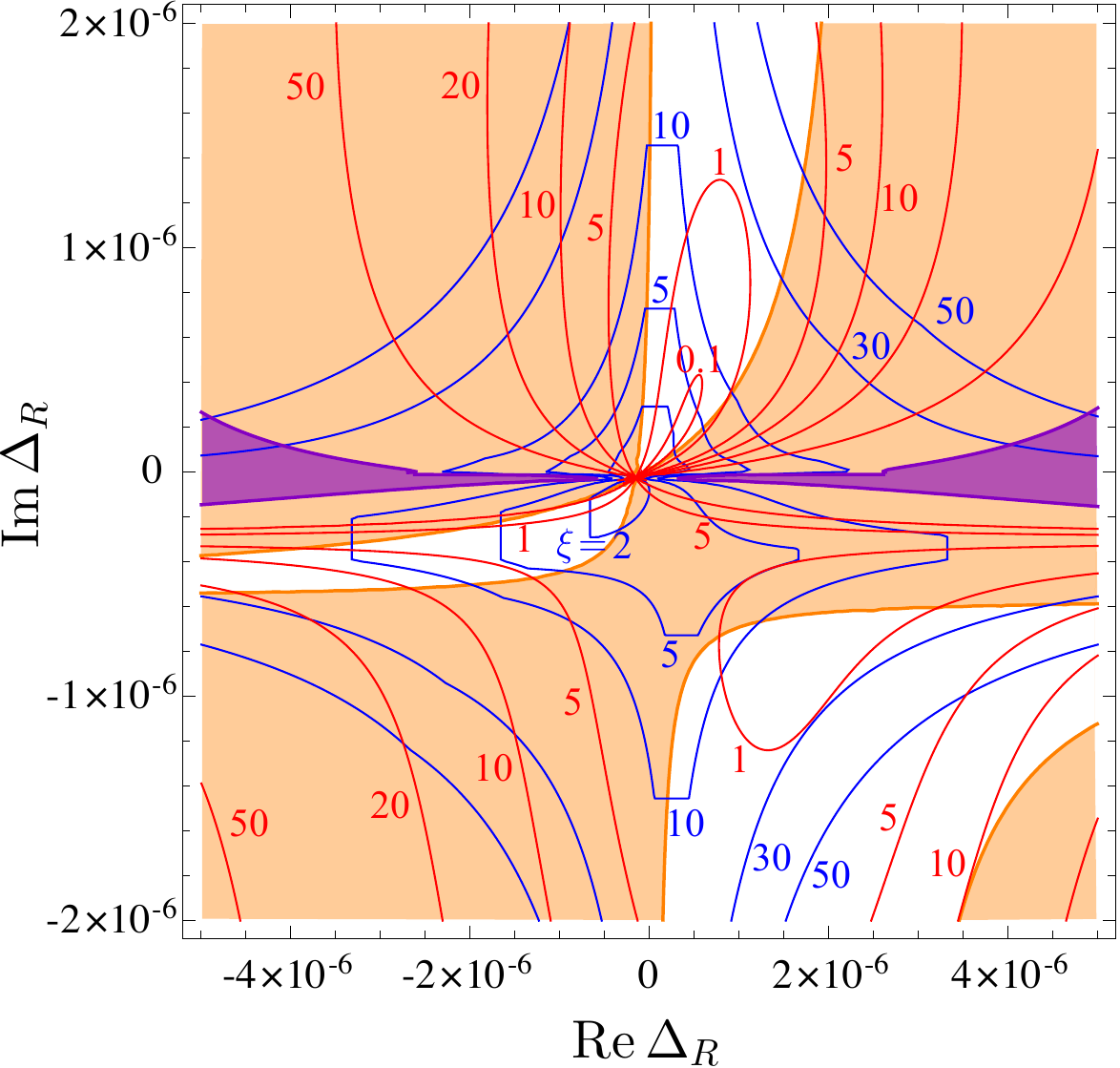}
}
~
\subfigure[$\mathcal{B}(K_L \to \pi^{0} \nu \bar{\nu})/\mathcal{B}(K_L \to \pi^{0} \nu \bar{\nu})_{\rm SM}$]
{
\includegraphics[width=0.465\textwidth, bb = 0 0 338 322]{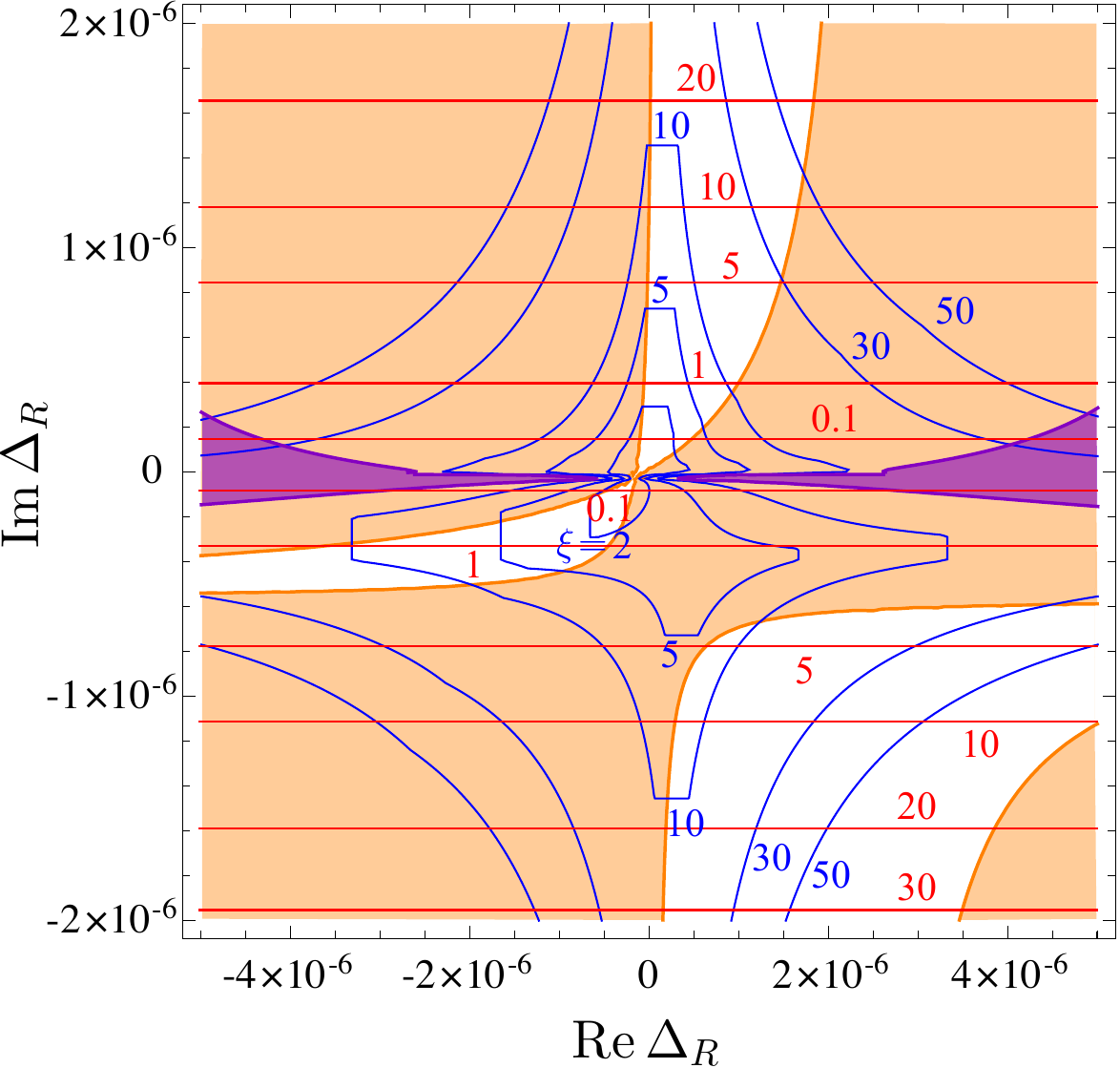}
}
\end{center}
\caption{
$\mathcal{B}(K^{+} \to \pi^{+} \nu \bar{\nu})/\mathcal{B}(K^{+} \to \pi^{+} \nu \bar{\nu})_{\rm SM}$ ({\em left} panel) and $\mathcal{B}(K_{L} \to \pi^{0} \nu \bar{\nu})/\mathcal{B}(K_{L} \to \pi^{0} \nu \bar{\nu})_{\rm SM}$ ({\em right}) are shown by the red contours. 
The blue contours represent the tuning parameter $\xi$.
The orange and purple shaded regions are excluded by $\mathcal{B}(K_L \to \mu^{+} \mu^{-})$ and $\Delta m_K$, respectively.
Here, $(\epsilon'/\epsilon )_{\rm NP} = 15.5 \cdot 10^{-4} $ and $\epsilon_K^{\rm NP} = 0.37 \cdot 10^{-3}$ as a reference.
The NP scale is set to be $\mu_{\rm NP} =1\TeV$.
}
\label{fig:generalZ}
\end{figure}

Let us consider the full parameter space in the general $Z$ scenario. 
Both $\Delta_{L}$ and $\Delta_{R}$ are turned on.
Then, $\mathcal{B}(K^{+}\to \pi^{+}\nu \bar{\nu})$ and/or $\mathcal{B}(K_{L}\to \pi^{0} \nu \bar{\nu} )$ can be enhanced if the tuning for $\epsilon_K^{\rm NP}$ is allowed.

In Fig.~\ref{fig:generalZ}, the branching ratios of $K \to \pi \nu \bar{\nu}$ and the tuning parameter are shown for the case of $(\epsilon'/\epsilon )_{\rm NP} = 15.5 \cdot 10^{-4} $ and $\epsilon_K^{\rm NP} = 0.37 \cdot 10^{-3}$.
The flavor-changing $Z$ couplings, and namely the NP contributions to $K \to \pi \nu \bar{\nu}$, are limited by $\mathcal{B}(K_L \to \mu^{+} \mu^{-})$ and the tuning parameter.

\begin{figure}[t]
\begin{center}
\includegraphics[width=0.7\textwidth, bb = 0 0 562 472]{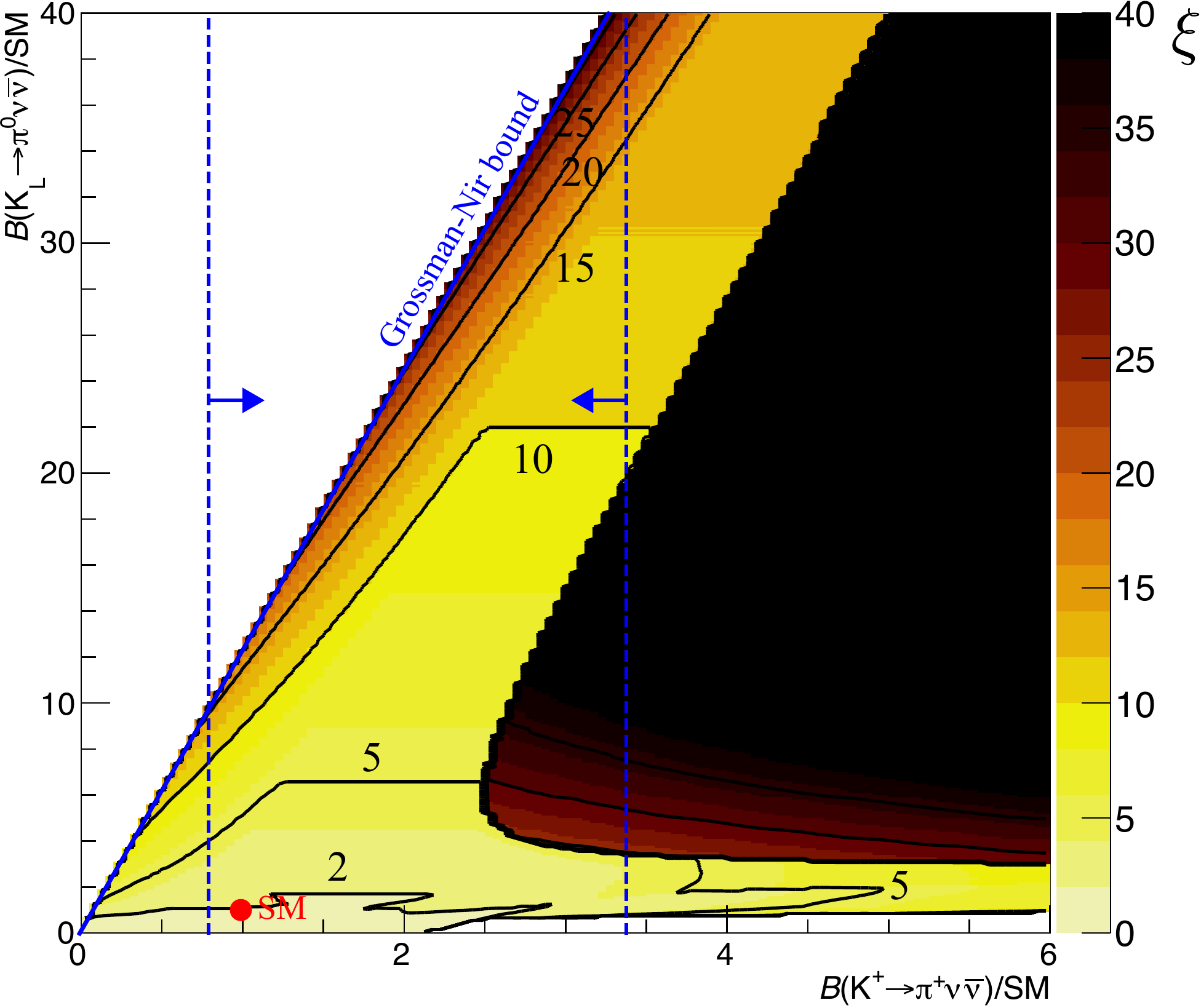}
\end{center}
\caption{
Contours of the tuning parameter $\xi$ are shown in the general $Z$ scenario.
In the colored regions, $\epsilon' / \epsilon$ is explained at $1\sigma$, and the experimental bounds of $\epsilon_K$, $\Delta m_K$, and $\mathcal{B}(K_L \to \mu^{+} \mu^{-})$ are satisfied.
The region between the blue dashed lines is allowed by the measurement of $\mathcal{B}(K^{+} \to \pi^{+} \nu \bar{\nu})$ at $1\sigma$.
There are no available model parameters above the Grossman-Nir bound.
The NP scale is set to be $\mu_{\rm NP} =1\TeV$.
}
\label{fig:generalZbranch}
\end{figure}

In Fig.~\ref{fig:generalZbranch}, contours of the tuning parameter $\xi$ are shown on the plane of the branching ratios of $K \to \pi \nu \bar{\nu}$.
The whole parameter space of the general $Z$ scenario is scanned.
In the colored regions, $\epsilon' / \epsilon$ is explained at the $1\sigma$ level, and the experimental bounds of $\epsilon_K$, $\Delta m_K$, and $\mathcal{B}(K_L \to \mu^{+} \mu^{-})$ are satisfied (see the previous section for the experimental constraints). 
For each set of $\mathcal{B}(K^{+}\to \pi^{+}\nu \bar{\nu})$ and $\mathcal{B}(K_{L}\to \pi^{0} \nu \bar{\nu} )$, the smallest $\xi$ is chosen among the parameter sets which predict the same branching ratios.

Compared to the simplified cases in Fig.~\ref{fig:simplified}, $\mathcal{B}(K_{ L}\to \pi^{0} \nu \bar{\nu} )$ can be enhanced.
The tuning parameter is not necessarily very large if only one of $\mathcal{B}(K_{L}\to \pi^{0} \nu \bar{\nu} )$ and $\mathcal{B}(K_L \to \mu^{+} \mu^{-})$ is enhanced.
However, $\xi \gtrsim 30$--$40$ is required to amplify both of them.
If $\xi \lesssim 10$ (5) is allowed, $\mathcal{B}(K_L \to \pi^{0} \nu \bar{\nu})$ can be as large as $6 \times 10^{-10}$ ($2 \times 10^{-10}$).
In other words, $\mathcal{O}(10)$\% tunings are required to enhance $\mathcal{B}(K_L \to \pi^{0} \nu \bar{\nu})$ by an order of magnitudes compared the SM prediction.
The KOTO experiment can probe such large branching ratios in the near future.

\section{Conclusion and discussion}
\setcounter{equation}{0}

The recent discrepancy of $\epsilon'/\epsilon$ may be a sign of the NP contribution to the flavor-changing $Z$  coupling. 
In this letter, we revisited the scenario with paying attention to the interference effects between the SM and NP contributions to the $\Delta S = 2$ observables. 
They affect $\epsilon_K$ significantly once the right-handed coupling is turned on.
Consequently, $\mathcal{B}(K_L \to \pi^{0} \nu \bar{\nu})$ is smaller than the SM prediction in the simplified scenarios as long as $\epsilon'/\epsilon$ is explained.

In the general $Z$ scenario, $\mathcal{B}(K_L \to \pi^{0} \nu \bar{\nu})$ can be large if parameter tunings are allowed. 
It was found that the branching ratio can be enhanced by an order of magnitudes compared to the SM prediction if the NP contributions to $\epsilon_K$ are tuned at the $\mathcal{O}(10)$\% level.
It can be as large as $6 \times 10^{-10}$ ($2 \times 10^{-10}$) for $\xi \simeq 10$ (5), which implies that the NP contributions to $\epsilon_K$ are tuned at the 10\,\% (20\,\%) level.
The KOTO experiment could probe such large branching ratios in the near future.

In the analysis, the NP scale was set to be $1\TeV$.
The NP contributions to $\epsilon_K$ as well as the tuning parameter depend on it through the interference terms of the SM and NP (see Eq.~\eqref{eq:NGboson}).
For $\mu_{\rm NP} \gtrsim 1\TeV$, $\Delta_L^{\rm SM}$ is enhanced as the NP scale increases.
Hence, one naively expects that tighter tuning is required in $\epsilon_K$.  
However, renormalization group corrections could be larger in such a case.
Such contributions will be studied in elsewhere (see also Ref.~\cite{Bobeth2017}).


\vspace{1em}
\noindent {\bf Note added}: 
While resubmitting the manuscript, Ref.~\cite{Bobeth2017} appeared on the arXiv. 
In comparison with our analysis, the differences are as follows:
\begin{itemize}
\item we considered $\mathcal{O}_L$ and $\mathcal{O}_R$ operators with the  first leading logarithmic renormalization group contribution $\ln (\mu_{\rm NP}/m_W) $ which comes from Fig.~\ref{fig:diagram}\,(e). 
Hence,  the operator $\mathcal{O}^{(3)}_{L}$, the operator mixing among them through the renormalization group above the electroweak scale, nor  running effects of the coupling constants are not considered in our analysis.
\item  a conservative constraint from $\mathcal{B}(K_L \to \mu^+\mu^-)$~\cite{Isidori:2003ts} is imposed here, while Ref.~\cite{Bobeth2017} has also adopted an aggressive bound~\cite{DAmbrosio:1997eof}.
\item the present analysis focuses on the parameter region where the $\epsilon' / \epsilon$ discrepancy can be explained at $1\,\sigma$ level.
\end{itemize}
The loop function $\widetilde C(x,\mu_{\rm NP})$ in Eq.~\eqref{eq:DeltaLSM}, which comes from  $\mathcal{O}_{L}$ and $\mathcal{O}_R$,  is in agreement with a result of Ref.~\cite{Bobeth2017}, which comes from  $\mathcal{O}_{Hq}^{(1)} $ and $\mathcal{O}_{Hd}$, at the first leading logarithmic approximation.  Notice that $\mu_{\rm NP}$ corresponds to $\mu_{\Lambda}$.

\vspace{1em}
\noindent {\it Acknowledgements}: 
We would like to thank  Andrzej J. Buras for useful discussions. 
We are indebted to the referee for important comments.
This work is supported by JSPS KAKENHI No.~16K17681 (M.E.) and 16H03991 (M.E.).


\end{document}